\documentclass[aps,prl,showpacs,groupedaddress,twocolumn,amsmath,amssymb]{revtex4-1}

\usepackage[american]{babel}
\usepackage{graphicx}
\usepackage{times}
\usepackage{color}
\usepackage{grffile}

\begin{document}


\title{Direct Sampling of Negative Quasiprobabilities of a Squeezed State}

\author{T. Kiesel and W. Vogel}

\affiliation{Arbeitsgruppe Quantenoptik, Institut f\"ur Physik, Universit\"at  Rostock, D-18051 Rostock,
Germany}

\author{B. Hage}
\affiliation{
ARC Centre of Excellence for Quantum-Atom Optics, Department of Quantum Science, The Australian National University, Canberra, Australian Capital Territory 0200, Australia
}

\author{R. Schnabel}

\affiliation{
Institut f\"{u}r Gravitationsphysik, Leibniz Universit\"{a}t Hannover
 and Max-Planck-Institut f\"{u}r Gravitationsphysik (Albert-Einstein-Institute),\\
Callinstrasse 38, 30167 Hannover, Germany
}

\begin{abstract}
Although squeezed states are nonclassical states, so far, their nonclassicality could not be demonstrated by negative quasiprobabilities. In this work we derive pattern functions for the direct experimental determination of so-called nonclassicality quasiprobabilities. The negativities of these quantities turn out to be necessary and sufficient for the nonclassicality of an arbitrary quantum state and are therefore suitable for a direct and general test of nonclassicality. We apply the method to a squeezed vacuum state of light that was generated by parametric down-conversion in a second-order nonlinear crystal.
\end{abstract}
\maketitle

\paragraph{Introduction.}
In quantum optics and quantum information science, the notion of nonclassicality describes the distinguished difference between classical and quantum physics. Here, a quantum state is referred to as nonclassical, if one is not able to model the outcomes of experimentally measured optical field correlation functions by classical electrodynamics. Considering solely pure states, the famous coherent states $\left|\alpha\right>$ are the only classical states according to this notation, which makes them the  closest analogue to the classical oscillator~\cite{naturw-14-664}. Sudarshan~\cite{prl-10-277} and Glauber~\cite{pr131-2766} showed that the density operator of an arbitrary quantum state can formally be written as a statistical mixture of coherent states, 
\begin{equation}
 	\hat\rho = \int d^2\alpha\, P(\alpha)\left|\alpha\right>\left<\alpha\right|.
\end{equation}
If $P(\alpha)$ resembles the properties of a classical probability function, the state is simply a classical mixture of the (classical) coherent states, e.g.~a thermal state. 
In general, the $P$ function may attain negative values -- often in connection with a strongly singular behavior. In such cases the corresponding quantum state is referred to as a nonclassical one~\cite{pr-140B-676}.

The main problem of this definition of nonclassicality lies in the singularities of the $P$ function, which definitely prevent the experimental reconstruction of $P(\alpha)$. Only for special quantum states one may approximately obtain this quasiprobability~\cite{pra-78-021804R}. Therefore, different criteria for the detection of nonclassicality have been developed. Some of them are simple, such as squeezing~\cite{nature-306-141}, classical limits on probabilities~\cite{pra-79-042105} or  negativities in the Wigner function~\cite{joptb-6-396}, but they are only sufficient for nonclassicality. Others are necessary and sufficient, but they consist of an infinite hierarchy of inequalities~\cite{prl-89-283601,pra-72-043808}. Recently, nonclassicality quasiprobabilities have been introduced, which provide a complete and simple method for the verification of nonclassicality~\cite{pra-82-032107}: For any nonclassical state, there exists a regular nonclassicality quasiprobability, whose negativities unambiguously reflect its nonclassicality.  In~\cite{Bellini}, the experimental applicability, as a matter of principle, was demonstrate on a nonclassical but less problematic state, which also had a negative Wigner function.

In this Letter, we prove the nonclassicality of a Gaussian squeezed state by reconstructing negative quasiprobabilities from data taken by a balanced homodyne detector. Squeezed states are of special interest in this context, since their commonly used quasiprobabilities, such as the Wigner function, are nonnegative and still satisfy the properties of classical probability densities.
We avoid any Fourier transformation of the data, which was used in \cite{Bellini}, and present a method of direct sampling of nonclassicality quasiprobabilities from the measured data. 
For this purpose, we use the concept of pattern functions~\cite{pra-50-4298}, which provide an estimate of the quasiprobability together with its variance. This method applies to the experimental characterization of nonclassicality of any quantum states, the only limitation being statistical uncertainties.

\paragraph{Quasiprobabilities of squeezed states.}
Squeezed states are prominent examples of nonclassical states, which can be easily generated in the laboratory. Although nonclassicality is defined by negativities of the $P$ function, its general verification 
by negativities of any commonly used quasiprobability distribution is impossible. For instance, the Wigner function of a squeezed state with quadrature variances $V_x$ and $V_p$ reads as
\begin{equation}
        W_{\rm sv}(x,p) = \frac{1}{2\pi\sqrt{V_xV_p}}
        \exp\left\{-\frac{x^2}{2V_x}-\frac{p^2}{2V_p}\right\},
	\label{eq:Wigner:squeezed}
\end{equation}
clearly being a Gaussian. In contrast, the $P$ function may formally be written as
 \begin{equation}
        P_{\rm sv}(\alpha) =
        e^{-\frac{V_x-V_p}{8}\left(\frac{\partial^2}{\partial
        \alpha^2} + \frac{\partial^2}{\partial \alpha^{*2}}  -
        2 \frac{V_x+V_p-2}{V_x-V_p}\frac{\partial}{\partial
        \alpha}\frac{\partial}{\partial \alpha^*}\right)}\delta(\alpha),
\end{equation}
representing one of the most singular representations of a quantum state, with infinitely high orders of derivatives of the $\delta$-distribution. Moreover, the $s$-parameterized quasiprobabilities~\cite{pr-177-1857} of a squeezed state are either Gaussian (and nonnegative) or strongly singular. Based on a simple condition for the characteristic function of the $P$~function~\cite{prl84-1849}, the nonclassicality can be easily verified~\cite{pra-79-022122}. However, this condition is sufficient only and the question remains if there exists any well-behaved quasiprobability, allowing a complete characterization of the nonclassicality of squeezed states by the failure of being interpreted as a classical probability.

\paragraph{Nonclassicality quasiprobabilities.}
The starting point of our discussion is the characteristic function of the $P$ function, 
\begin{equation}
 	\Phi(\beta) = \langle :e^{\beta\hat a^\dagger - \beta^*\hat a}:\rangle = \langle e^{i |\beta| \hat x[{\rm arg}(\beta)-\pi/2]}\rangle e^{|\beta|^2/2},\label{eq:def:Phi}
\end{equation}
with $\hat x(\varphi)$ being the quadrature operator of the optical field at phase $\varphi$. In order to obtain a regular phase-space distribution, we filter $\Phi(\beta)$ in the form 
\begin{equation}
 	\Phi_\Omega(\beta) = \Phi(\beta)\Omega_w(\beta). \label{eq:def:Phi:Omega}
\end{equation}
The filter $\Omega_w(\beta)$ has to satisfy the following conditions to be useful for nonclassicality detection~\cite{pra-82-032107}:
\begin{enumerate}
 	\item The filtered characteristic function $\Phi_\Omega(\beta)$ should be integrable for an arbitrary quantum state $\Phi(\beta)$, such that its Fourier transform -- the nonclassicality quasiprobability -- exists as a regular function.
	\item Negativities in the Fourier transform shall unambiguously be due to the nonclassicality of the state. Conversely, the nonclassicality quasiprobability shall be nonnegative for any classical state. This requires that the filter $\Omega_w(\beta)$ has a nonnegative Fourier transform.
	\item If the width parameter approaches infinity, the filtered characteristic function $\Phi_\Omega(\beta)$ should converge to the characteristic function of the $P$~function, $\Phi(\beta)$. This can be realized by the conditions $\Omega_w(\beta) = \Omega_1(\beta/w)$ and $\Omega_1(0) = 1$.
	\item The filter should be nonzero everywhere, $\Omega_w(\beta) \neq 0$, such that no information about the quantum state is lost due to the filtering in Eq.~(\ref{eq:def:Phi:Omega}).
\end{enumerate}

Under these conditions, the nonclassicality quasiprobability is defined as the Fourier transform of the filtered characteristic function, 
\begin{equation}
 	P_\Omega(\alpha) = \frac{1}{\pi^2}\int d^2\beta e^{\alpha\beta^*-\alpha^*\beta} \Phi(\beta) \Omega_w(\beta) \label{eq:Phi:to:P}.
\end{equation}
Negativities of the quasiprobability are signatures of nonclassicality of the state and the negativity of its $P$ function, since the filter is constructed in such a way that it does not introduce negativities by itself.
In the present work, we construct a filter from the autocorrelation of the function $\omega(\beta) = \exp(-|\beta|^4)$,
\begin{equation}
	\Omega_1(\beta) = \frac{1}{\mathcal N}\int d^2\beta' \omega(\beta')\omega(\beta'+\beta),
\end{equation}
the normalization constant ${\mathcal N}$ is chosen to obey $\Omega_1(0) = 1$. The width is introduced via $\Omega_w(\beta) = \Omega_1(\beta/w)$. This filter satisfies all criteria mentioned above, for the proof see~\cite{pra-82-032107}.

\paragraph{Derivation of a pattern function.}
Pattern functions provide an efficient technique to directly estimate a physical quantity together with its uncertainty.  From balanced homodyne detection, we obtain quadrature values $x_j(\varphi)$ measured for certain phases $\varphi$. They obey the quadrature distributions $p(x; \varphi)$, which satisfy $\int p(x;\varphi) dx = 1$. The quadratures are normalized such that the vacuum quadratures have a variance $V_{\rm vac} = 1$. Now the nonclassicality quasiprobability $P_\Omega(\alpha)$ with a certain width parameter $w$ is written as the statistical mean of the pattern function $f_\Omega(x,\varphi; \alpha,w)$, averaged over the quadrature distributions:
\begin{equation}
  P_\Omega(\alpha) = \int_{-\infty}^\infty dx \int_0^\pi d\varphi \, \frac{p(x; \varphi)}{\pi} f_\Omega(x, \varphi; \alpha, w).\label{eq:expect:of:pattern:f}
\end{equation}
For this purpose, we note that due to Eq.~(\ref{eq:def:Phi}), the characteristic function of the $P$ function of the state can be calculated from the quadrature distribution via
\begin{equation}
 	\Phi(\beta) = \int_{-\infty}^\infty dx\, p\left(x;\arg(\beta)-\tfrac{\pi}{2}\right) e^{i|\beta| x} e^{|\beta|^2/2}. \label{eq:p(x):to:Phi}
\end{equation}
It is convenient to rewrite the integral in Eq.~(\ref{eq:Phi:to:P}) in polar coordinates $\beta = b e^{i\varphi}$. Here, we restrict $\varphi$ to $[0,\pi)$ and extend $b$ to $(-\infty,\infty)$. Then we obtain
\begin{align}
 	P_\Omega(\alpha) =& \frac{1}{\pi^2}\int_{-\infty}^\infty db \int_0^\pi d\varphi |b| e^{2i|\alpha| b \sin(\arg(\alpha)-\varphi)}  \Phi(b e^{i\varphi}) \nonumber\\
		\times&\Omega_w(b e^{i\varphi}).
\end{align}
The filter is chosen to be independent of the phase, i.e.~$\Omega_w(b e^{i\varphi}) \equiv \Omega_w(b)$. Now we insert Eq.~(\ref{eq:p(x):to:Phi}) and obtain
\begin{align}
P_\Omega(\alpha) =&\int_{-\infty}^\infty dx \int_0^\pi d\varphi \, \frac{p(x; \varphi)}{\pi}\int_{-\infty}^\infty  db \, \frac{|b|}{\pi} e^{i b x}  e^{b^2/2} \nonumber\\
\times&e^{2i|\alpha| b \sin(\arg(\alpha)-\varphi-\tfrac{\pi}{2})} \Omega_w(b).
\label{eq:sampling:P}
\end{align}
This equation defines the pattern function 
\begin{align}
	f_\Omega(x, \varphi; \alpha, w) =& \int_{-\infty}^\infty db\,\frac{|b|}{\pi} e^{i b x}  e^{2i|\alpha| b \sin(\arg(\alpha)-\varphi-\tfrac{\pi}{2})} \nonumber\\ \times& e^{b^2/2}\Omega_w(b),  \label{eq:def:pattern:f}
\end{align}
which has to be used in Eq.~(\ref{eq:expect:of:pattern:f}).

Equation~(\ref{eq:expect:of:pattern:f}) gives rise to the following interpretation: Suppose we have measured $N$ quadrature-phase pairs $(x_i, \varphi_i)$, whose joint probability distribution is $\tfrac{1}{\pi}p(x;\varphi)$. Here the phases $\varphi$ are assumed to be uniformly distributed in $[0,\pi)$, while the quadratures obey the quadrature distribution $p(x;\varphi)$, conditioned on the value of the phase $\varphi$. Then the quasiprobability $P_\Omega(\alpha)$ can be calculated as the expectation value of the pattern function $f_\Omega(x, \varphi; \alpha, w)$. For experimental data, we replace the expectation value by its empirical estimate, 
\begin{equation}
P_\Omega(\alpha) \approx \frac{1}{N} \sum_{i=1}^N f_\Omega(x_i, \varphi_i; \alpha, w).
\end{equation}
Its variance can be obtained naturally as the mean square deviation of the numbers $f_\Omega(x_i, \varphi_i; \alpha, w)$.

If the phases, at which quadratures are measured, are scanned in $[0,\pi]$ or drawn randomly from a uniform distribution, one can calculate the nonclassicality quasiprobability directly as the empirical mean of the pattern function. This mean is taken over all pairs $(x_i,\varphi_i)$ of quadrature and phase. In our experiment, we only obtained quadratures at $21$ fixed phase angles. In this case, one may not simply replace the integral over the phase $\varphi$ by a sum:  This leads to systematic deviations, since the pattern function is varying rapidly with respect to the phase, in particular if $|\alpha|$ becomes large. For a detailed discussion and solution of this problem, we refer to the supplemental material~\cite{Supp}. 

\paragraph{Experimental set-up.}

The squeezed vacuum states of light were generated by type-I degenerate parametric down-conversion (optical parametric amplification, OPA) inside an optical resonator. The latter was a standing wave resonator with a line width of 25$\,$MHz containing of a non-critically phase matched second-order nonlinear crystal (7\% Mg:LiNbO$_3$). The OPA process was continuously pumped by 50$\,$mW of second harmonic light yielding a classical power amplification factor of six. Both, the length (resonance frequency) of the resonator as well as the orientation of the squeezing ellipse were stably controlled by electronic servo loops. With this setup we directly measured a squeezed variance of -4.5$\,$dB and an anti-squeezed variance of +7.2$\,$dB with respect to the unity vacuum variance. From these measurements we inferred an overall efficiency of 75\% and an initial squeezing factor of -8.2$\,$dB.

\begin{figure}[h]
\includegraphics[width=\linewidth]{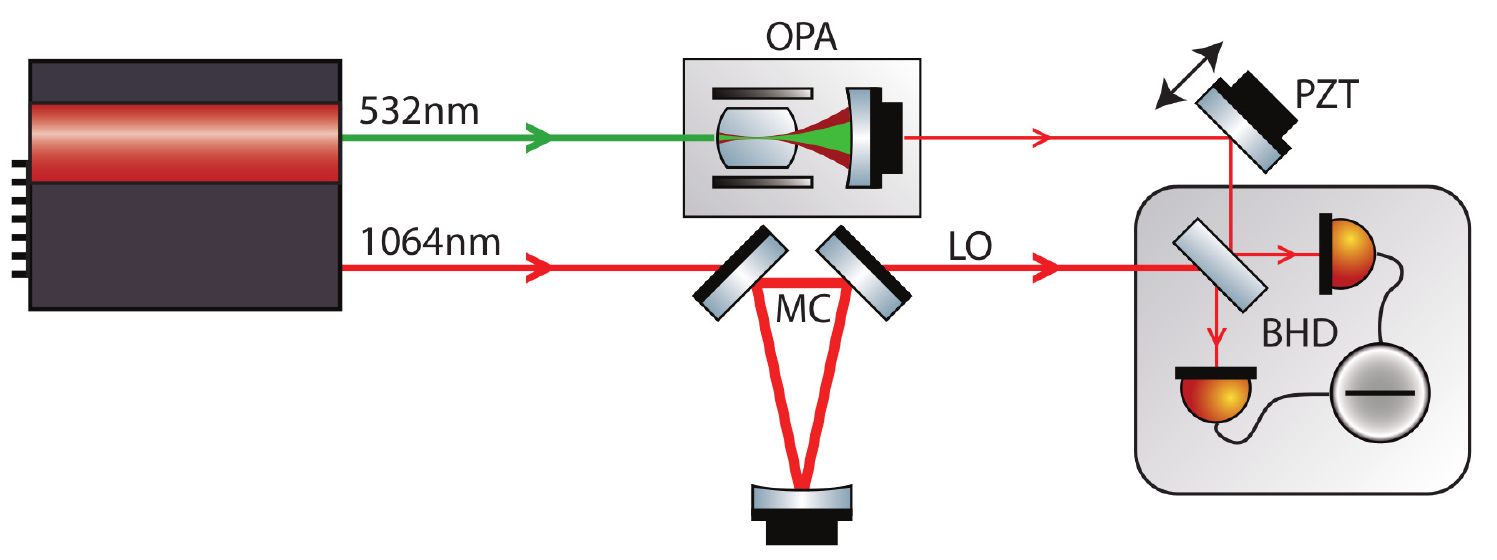}
\caption{Simplified sketch of the experimental setup. A spatially filtered continuous-wave field at 1064\,nm served as a local oscillator (LO) for balanced homodyne detection (BHD) and a phase-locked second harmonic field at 532\,nm as the pump for the parametric squeezed light source (OPA). MC: spatial mode cleaner, PZT: piezo-electrically actuated mirror for adjusting the quadrature amplitude phase of the BHD.
}
\label{fig:setup}
\end{figure}

The  quadrature amplitudes of the squeezed state were measured by balanced homodyne detection (BHD). The visibility of the squeezed field and the spatially filtered (MC, Fig.~\ref{fig:setup}) local oscillator was 98.9\% and was limited by OPA crystal inhomogeneities. The quadrature phase of the BHD was adjusted by servo loop controlled micro-positioning of steering mirrors in one of the optical input paths. The photo-electric signals of the two individual BHD-photodiodes were electronically mixed down at 7$\,$MHz and low pass filtered with a bandwidth of 400$\,$kHz to address a mode showing good squeezing and a high detector dark noise clearance of the order of 20$\,$dB. The resulting signals were fed into a PC based data acquisition system and sampled with one million samples per second and 14 bit resolution and finally subtracted yielding the quadrature amplitude data.  For a more detailed description of the main parts of the setup we refer to 
\cite{HageSQZpuri,FranzenSQZpuri}.

\begin{figure}
 	\includegraphics[width=\columnwidth]{{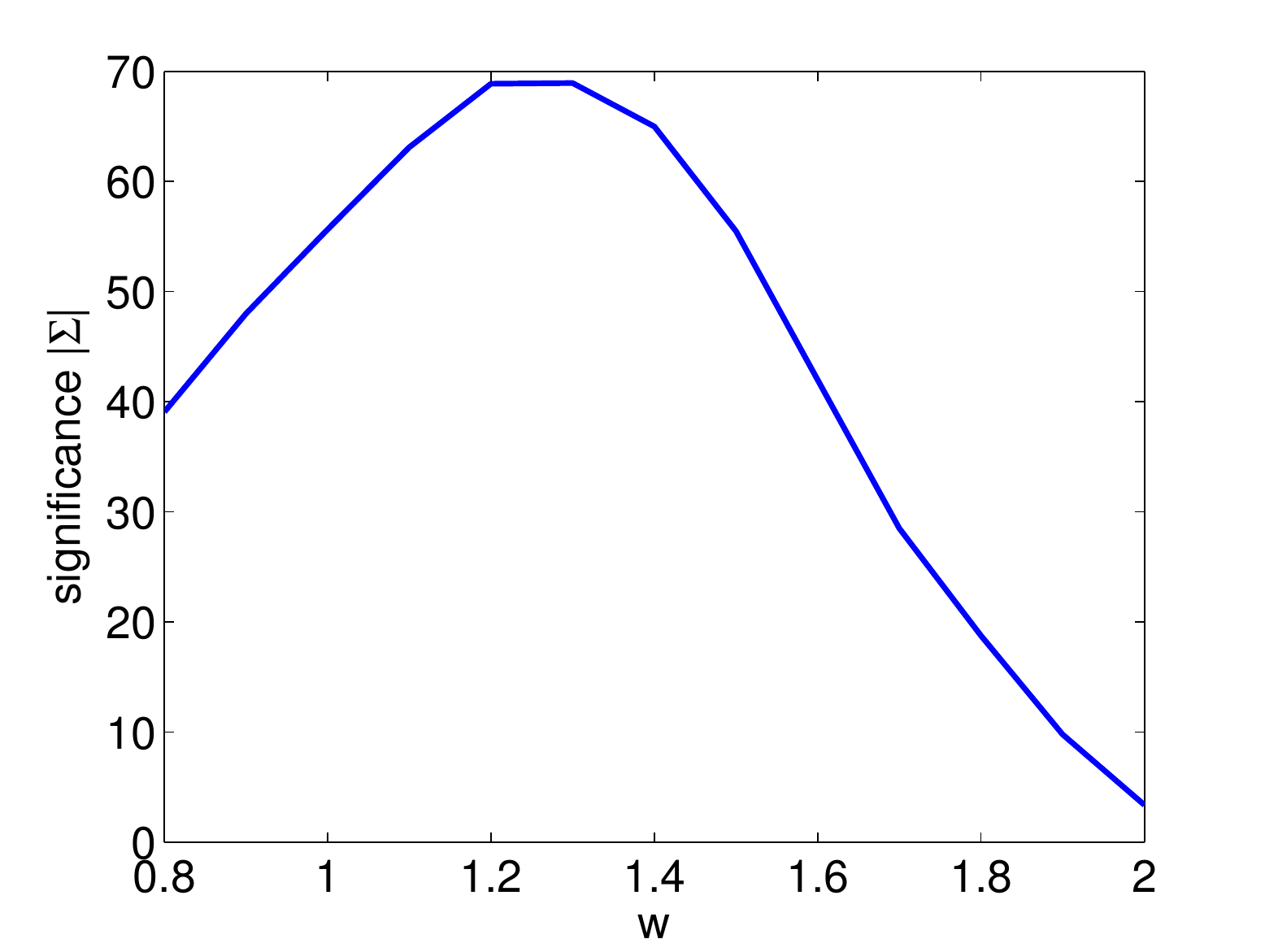}}
	\caption{Absolute value of the significance $\Sigma(w)$ of the negativity of the quasiprobability versus the filter width $w$.}
	\label{fig:opt:w}
\end{figure}

\paragraph{Experimental results.}
The examined squeezed vacuum state is characterized by the variances $V_x = 0.36$ and $V_p = 5.28$. We acquired $10^5$ quadrature values for each of the $21$ quadrature phases, the latter being equally spaced in $[0,\pi]$. The values of the quasiprobability $P_\Omega(\alpha)$ as well as their standard deviation $\sigma(P_\Omega(\alpha))$ are estimated from the pattern function as given in Eq.~(\ref{eq:def:pattern:f}). The filter width is chosen such that the significance of the negativity is optimized. Our figure of merit is defined as
\begin{equation}
	\Sigma(w) = \min_\alpha\left[ \frac{P_\Omega(\alpha)}{\sigma(P_\Omega(\alpha))}\right],
\end{equation}
with $\Sigma(w)$ being negative if $P_\Omega(\alpha)$ is negative for some $\alpha$. By construction of the quasiprobability, a significant negativity clearly indicates nonclassicality of the state. The larger the absolute value of $\Sigma(w)$, the larger is the significance of the negativity. This quantity can be optimized with respect to $w$. In Fig~\ref{fig:opt:w}, we show the dependence of the significance on the filter width $w$. The larger the filter width, the larger the nonclassical effects of the state are manifested in negativities, but the larger also the statistical uncertainty grows. Therefore, an optimum width exists, which is achieved here for $w = 1.3$.

\begin{figure}
 	\includegraphics[width=\columnwidth]{{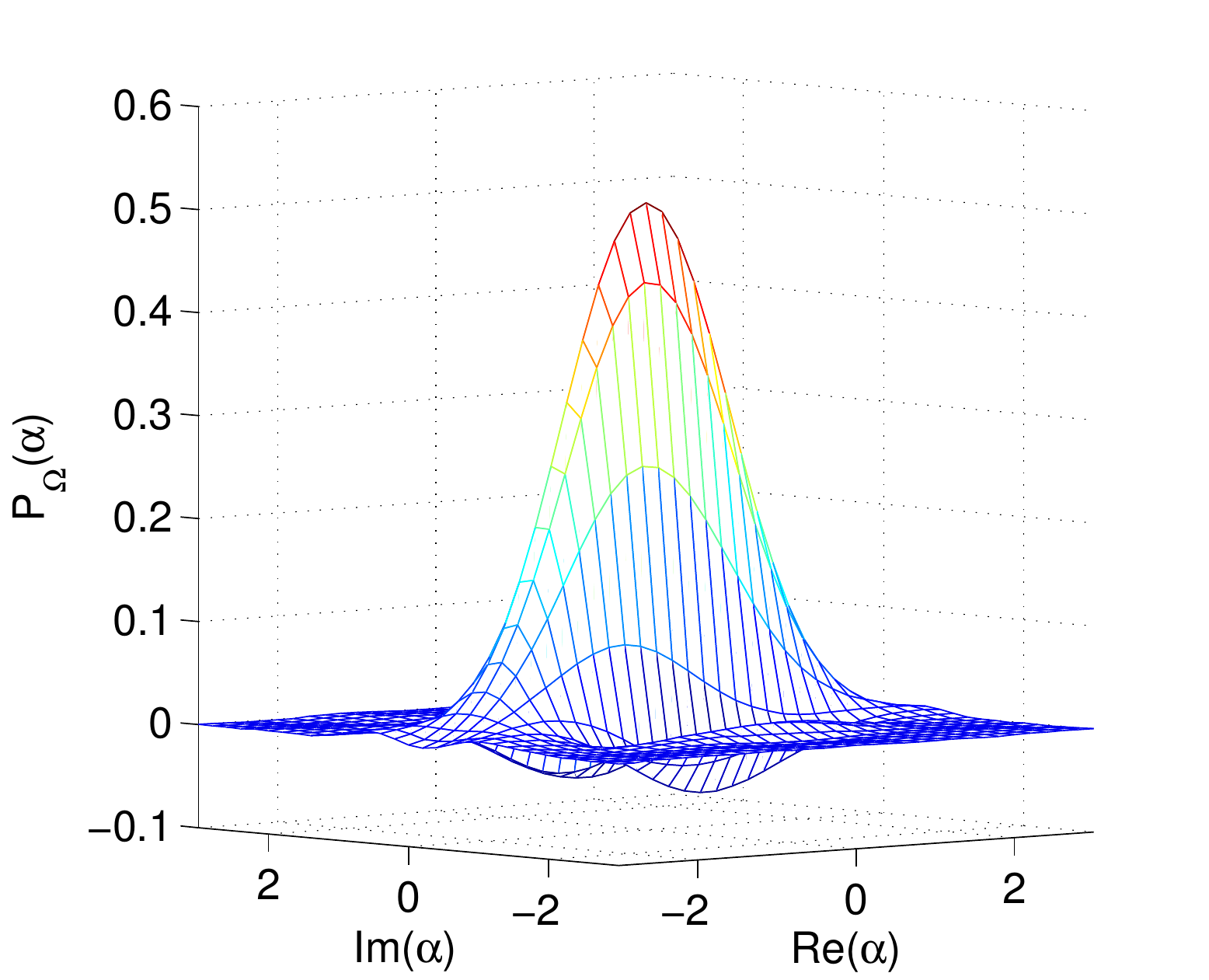}}
	\caption{Nonclassicality quasiprobability for a squeezed vacuum state. The data is directly sampled from our balanced homodyne data and clearly shows negative values.}
	\label{fig:P:Omega:2D}
\end{figure}

\begin{figure}
 	\includegraphics[width=\columnwidth]{{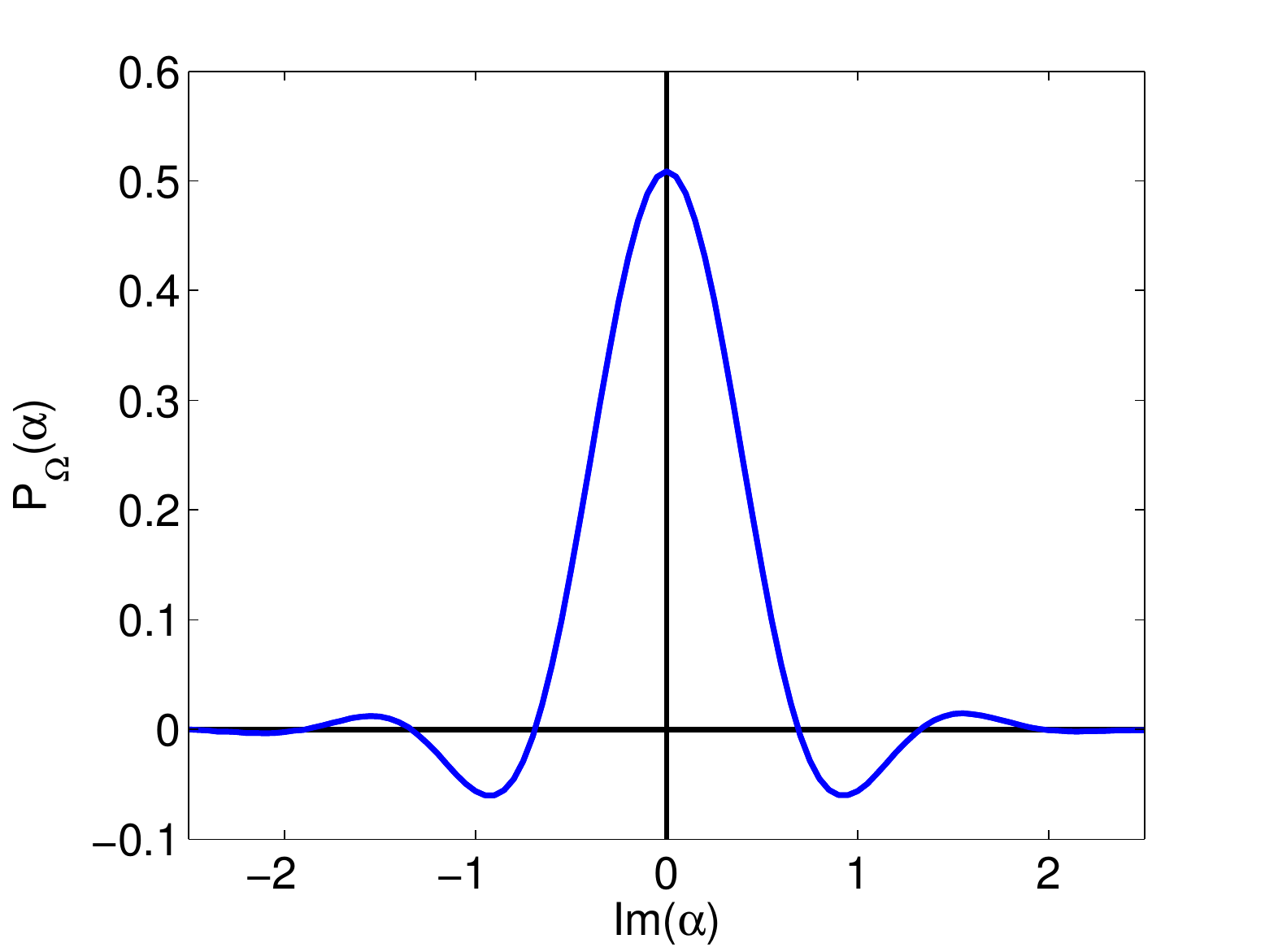}}
	\caption{Cross section of the nonclassicality quasiprobability for our squeezed vacuum state. Note, that the uncertainty in the data is less than the line width chosen here.}
	\label{fig:P:Omega:cross}
\end{figure}

Figure~\ref{fig:P:Omega:2D} shows the nonclassicality quasiprobability obtained from the experimental data. We find that along the axis of ${\rm Im}(\alpha)$, the quasiprobability oscillates and becomes clearly negative. This uncovers the  nonclassicality of the squeezed state in a general sense, beyond
the phase-sensitive reduction of the quadrature variance. 
It also includes the information on other kinds of effects, such as higher-order squeezing of the types considered in~\cite{pra-72-043808} and \cite{prl54-974}, and others, see~\cite{pra-79-022122}. In Fig.~\ref{fig:P:Omega:cross}, we show a cross-section of Fig.~\ref{fig:P:Omega:2D} along this axis. We clearly observe distinct negativities. The standard deviation is less than $1.1\times 10^{-3}$ for all points and therefore covered by the width of the line. We also calculated a systematic error due to the finite set of examined phases, being less than $3.6\times 10^{-4}$ for all points along this axis~\cite{Supp}.  $P_\Omega(\alpha)$ attains the minimum at $\alpha = 0.9 i$ with $P_\Omega(\alpha_{\min}) = -0.05989$ and $\sigma(P_\Omega(\alpha_{\min})) = 0.9\times 10^{-3}$, therefore leading to a significance of $|\Sigma| = 69$ standard deviations. Hence, this is a very clear demonstration of the nonclassicality of the examined state by means of negativities of a nonclassicality quasiprobability, which is not possible for commonly used quasiprobabilities such as the Wigner function.

\paragraph{Conclusions.}
In our work, we introduced a method for the direct sampling of nonclassicality quasiprobabilities of arbitrary quantum states from measured quadrature amplitudes. By applying our method to a squeezed state, whose $P$~function belongs to the most singular ones, we experimentally verified nonclassicality in its general sense, i.e. through negative quasiprobabilities. Our method is not only capable of estimating  the negativity of the quasiprobability, but also its statistical uncertainty, in a surprisingly simple manner. 

\paragraph{Acknowledgments.}
This work was supported by the Deutsche Forschungsgemeinschaft through SFB 652 and the Centre for Quantum Engineering and Space-Time Research, QUEST.

\newpage

\begin{center}
	\bf Supplementary Material
\end{center}

\paragraph{Numerical evaluation of the pattern function.}

The pattern function is given by
\begin{align}
	f_\Omega(x, \varphi; \alpha, w) =& \int_{-\infty}^\infty db\, \frac{|b|}{\pi} e^{i b x}  e^{2i|\alpha| b \sin(\arg(\alpha)-\varphi-\tfrac{\pi}{2})} \nonumber\\ \times& e^{b^2/2}\Omega_w(b).  \label{eq:def:pattern:f}
\end{align}
Since it has to be calculated for a large set of data points, its fast evaluation is of great relevance. First, we note that it can be regarded as a function of two parameters, 
\begin{equation}
  \chi(\xi; w) = \int_{-\infty}^\infty db \frac{|b|}{\pi} e^{i b \xi} e^{b^2/2}  \Omega_w(b)  , \label{eq:phi:as:four:integral}
\end{equation}
which is connected to $f_\Omega(x, \varphi; \alpha, w)$ via
\begin{equation}
 	f_\Omega(x, \varphi; \alpha, w) \equiv \chi(x + 2|\alpha|\sin(\arg(\alpha)-\varphi-\pi/2); w).
\end{equation}
 The data points only enter in $\chi$ via the argument $\xi$, while $w$ is arbitrary but fixed for all data points. Therefore, it is convenient to determine the function $\chi(\xi; w)$ in advance. This can be done by Fourier techniques: Since $\Omega_w(b)$ decays very fast with $b$, we may approximate it by setting it to zero for all $b$ with $|b| > b_c$, with some $b_c$ being sufficiently large. Note that $b_c$ scales with the width $w$, since $\Omega_w(\beta)$ is just a fixed function $\Omega_1(\beta)$ with a scaled argument $\beta\to\beta/w$. With this assumption, $\chi(\xi;w)$ is a bandlimited function, i.e.~its Fourier transform has bounded support. Therefore, the Nyquist-Shannon sampling theorem holds~\cite{Jerri}:
\begin{equation}
 	\chi(\xi;w) = \sum_{j = -\infty}^\infty \chi(\tfrac{\pi j}{b_c} ; w) \frac{\sin(b_c \xi - \pi j)}{b_c \xi - \pi j}.\label{eq:Nyquist}
\end{equation}
Hence, the Fourier integral~(\ref{eq:phi:as:four:integral}) has to be evaluated at a discrete set of points $\xi_j = \tfrac{\pi j}{b_c}$. Afterwards, the data points are inserted via $\xi$ in Eq.~(\ref{eq:Nyquist}). In our calculations, the support of $\Omega_w(\beta)$ was taken to be $b_c = 4 w$, where the filter $\Omega_w(b_c)$ became less than $10^{-15}$. Furthermore, we evaluated $256$ coefficients $\chi(\tfrac{\pi j}{b_c} ; w)$ for a numerical accuracy of $3.5\times 10^{-5}$ of $\chi(\xi;w)$.

\paragraph{Sampling from a discrete set of phases.}
The points of the nonclassicality quasiprobability can be obtained from the quadrature distributions by
\begin{equation}
		P_\Omega(\alpha) = \int_{-\infty}^\infty\int_0^\pi \frac{p(x; \varphi)}{\pi}f(x,\varphi; \alpha, w) d\varphi dx.\label{eq:sampling:P}
\end{equation}
Let us now consider the case that one has measured quadrature distributions at a certain set of phases $\{\varphi_k\}_{k=1}^{N_\varphi}$. 

As a first idea, one might replace the integral over the phase $\varphi$ by a Riemannian sum: 
\begin{equation}
	P_\Omega(\alpha) \approx \frac{1}{N_\varphi}\sum_{k=1}^{N_\varphi}\int_{-\infty}^\infty p(x; \varphi_k) f_\Omega(x,\varphi_k; \alpha, w)dx. \label{eq:P_Omega:Riemann}
\end{equation}
However, this leads to significant systematic deviations, such that $P_\Omega$ does not appear to be integrable. This can be seen as follows: The function $\chi(\xi;  w)$ is the Fourier transform of some integrable function, see Eq.~(\ref{eq:phi:as:four:integral}). Due to the Riemann-Lebesque lemma~\cite{Bochner:Buch}, when $\xi$ approaches infinity, $\chi(\xi; w)$ tends to zero. In terms of $\alpha$, this can be achieved if $|\alpha|\to\infty$, but $\arg(\alpha)-\varphi_k \neq \pi/2$ for all $\varphi_k$. However, what happens if the latter condition is not satisfied? Then, $f_\Omega(x,\varphi_k; \alpha, w)$ appears to be independent of $|\alpha|$ and does not approach zero for large $\alpha$, and the same holds for $P_\Omega(\alpha)$. Therefore, when one reconstructs $P_\Omega(\alpha)$ via Eq.~(\ref{eq:P_Omega:Riemann}), for all arguments of $\alpha$, which match one of the phases examined in experiment, the result does not approach zero but tends to a finite value. 
\begin{figure}[h]
 	\includegraphics[width=\columnwidth]{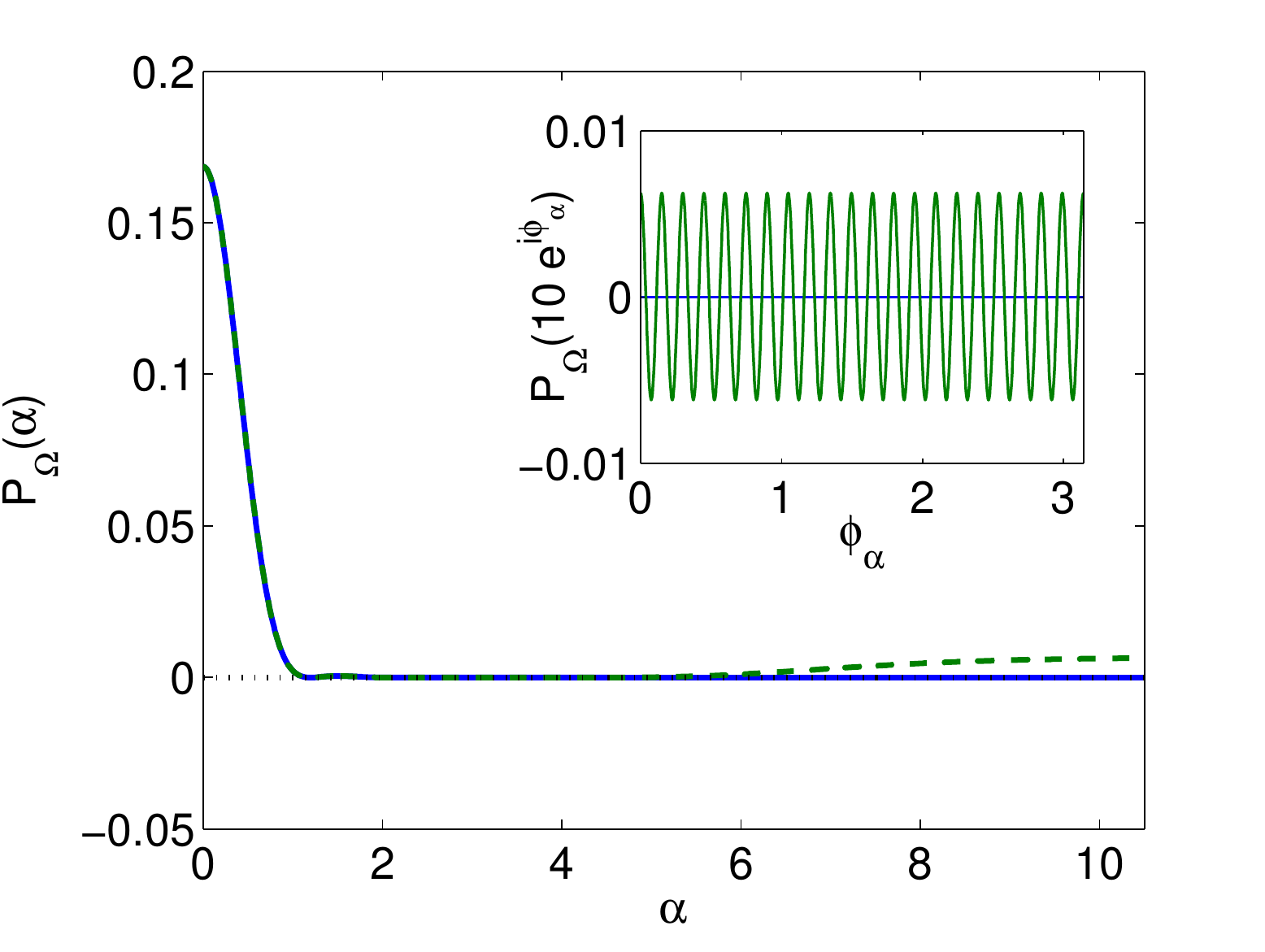}
	\caption{Application of the reconstruction formulae~\mbox{(\ref{eq:sampling:P})} (solid line) and (\ref{eq:P_Omega:Riemann}) (dashed line) for a vacuum state. It is clearly seen that the latter graph does not tend to zero for large $\alpha$. The inset shows the dependence of $P_\Omega(\beta)$ on the phase, with $|\alpha| = 10$.}
	\label{fig:Riemann:sum:vac}
\end{figure}

This effect can  already be observed for phase-independent states, such as vacuum. Figure~\ref{fig:Riemann:sum:vac} shows the exact curve (\ref{eq:sampling:P}) (solid line) and the approximation~(\ref{eq:P_Omega:Riemann}) (dashed line) for a vacuum state. We assume that the state has been examined at $21$ phases, and plot the cross section of $P_\Omega$ along one of them. It is clearly seen that the approximation~(\ref{eq:P_Omega:Riemann}) does not tend to zero for large $\alpha$. The inset shows the dependence on the phase of $\alpha$, for $|\alpha| = 10$. We observe an oscillating function with $21$ maxima, which do not vanish for larger $\alpha$. This artefact is caused by the fact that the pattern function $f_\Omega(x,\varphi; \alpha, w)$ strongly oscillates with $\varphi$ when $\alpha$ increases. Therefore, the phase integral in Eq.~(\ref{eq:sampling:P}) cannot be properly approximated by a sum. 

\begin{figure}
	\includegraphics[width=\columnwidth]{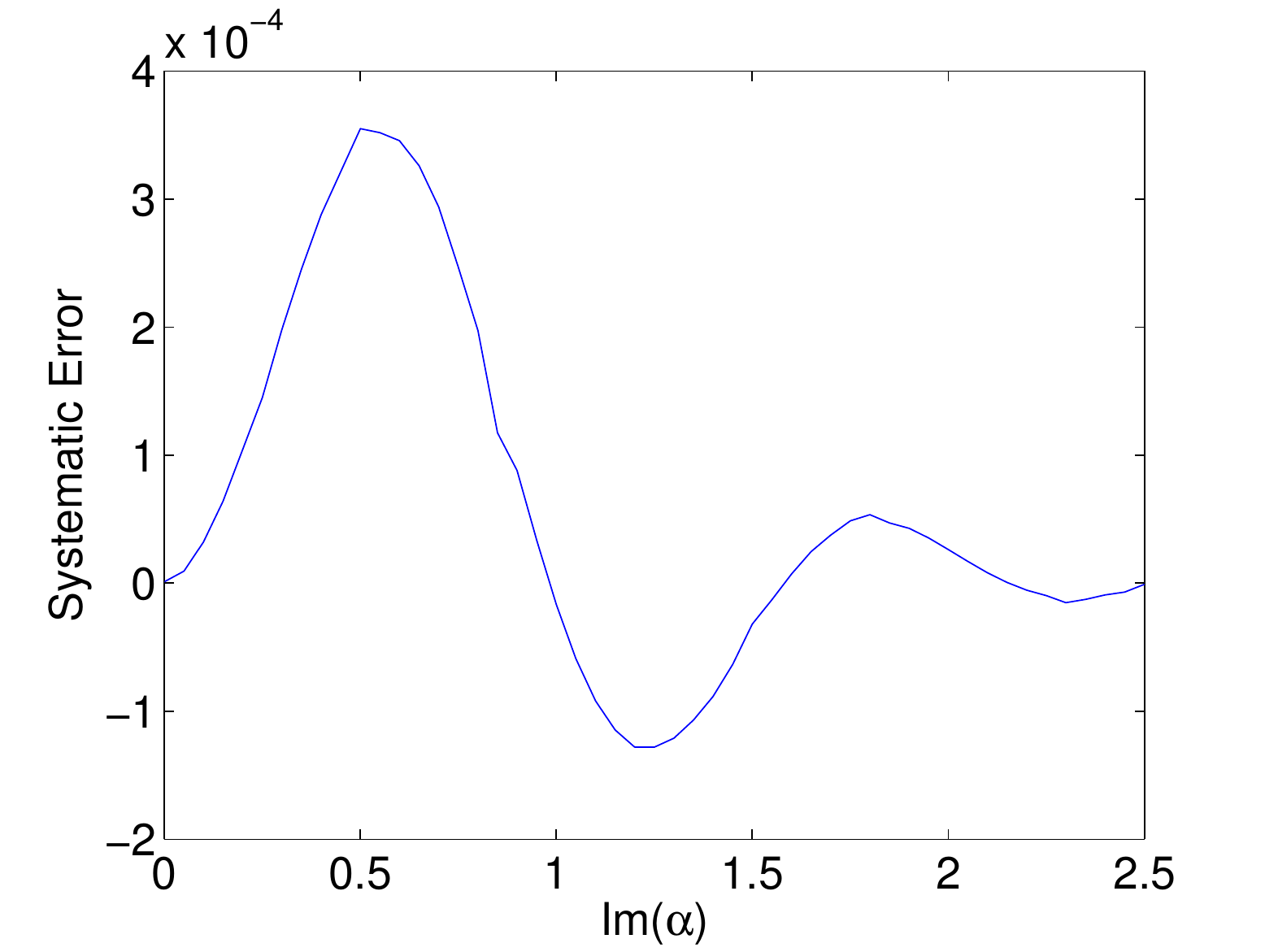}
	\caption{Systematic error due to a discrete set of examined phases.}
	\label{fig:syst:err:cross}
\end{figure}

In the following, we assume that only the quadrature distributions $p(x; \varphi)$ do not change significantly within the intervals $[\varphi_k - \tfrac{\pi}{2N_\varphi}, \varphi_k + \tfrac{\pi}{2N_\varphi}]$ for all $k = 1,\ldots,N_\varphi$. More illustratively, we assume that our set of quadrature distributions measured at the phases $\varphi_k$ suffices to have complete information about the quadrature distributions at phases $\varphi$ which have not been measured. If this were not the case, one had to increase the phase resolution in order to obtain more information. With this assumption, we rewrite Eq.~(\ref{eq:sampling:P}) in the following way:
\begin{align}
	P_\Omega(\alpha) &= \int_{-\infty}^\infty\int_0^\pi \frac{p(x; \varphi)}{\pi} f(x,\varphi;\alpha,w) d\varphi dx\nonumber\\
	=&\sum_{k=1}^{N_\varphi}\int\int_{-\tfrac{\pi}{2N_\varphi}}^{\tfrac{\pi}{2N_\varphi}} \frac{p(x; \varphi_k+\varphi)}{\pi} f(x,\varphi_k+\varphi;\alpha,w) d\varphi dx\nonumber\\
	\approx &\sum_{k=1}^{N_\varphi}\int\frac{p(x; \varphi_k)}{\pi}\int_{-\tfrac{\pi}{2N_\varphi}}^{\tfrac{\pi}{2N_\varphi}}  f(x,\varphi_k+\varphi;\alpha,w) d\varphi dx\nonumber\\
	=&\frac{1}{N_\varphi}\sum_{k=1}^{N_\varphi}\int p(x; \varphi_k)\tilde f(x,\varphi_k;\alpha,w) dx\label{eq:P:Omega:approx},
\end{align}
where the modified pattern function is defined as
\begin{equation}
	\tilde f(x,\varphi_k;\alpha,w) = \frac{N_\varphi}{\pi}\int_{-\tfrac{\pi}{2N_\varphi}}^{\tfrac{\pi}{2N_\varphi}}  f(x,\varphi_k+\varphi;\alpha,w) d\varphi.\label{eq:pattern:f:mod}
\end{equation}
The difference between the first and the last line in Eq.~(\ref{eq:P:Omega:approx}) is a systematic error, caused by measuring at a discrete set of phases. Figure~\ref{fig:syst:err:cross} shows this quantity for a squeezed vacuum state with the measured variances $V_x = 0.36$ and $V_p = 5.28$ along the axis where negativities of the nonclassicality quasiprobability are found. There, the systematic deviations are approximately one order of magnitude below the statistical standard deviation, which has been given in the Letter.

The modified pattern function (\ref{eq:pattern:f:mod}) can be interpreted as the original pattern function, averaged over the phase $\varphi$, which is uniformly distributed over the interval $[-\tfrac{\pi}{2N_\varphi}, \tfrac{\pi}{2N_\varphi}]$. We realize this by drawing a random number $\varphi$ in this interval for each quadrature point $x_j(\varphi_k)$, which has been measured at phase $\varphi_k$, and add this random number $\varphi$ to the phase $\varphi_k$. In this way, $\varphi_k$ represents a random phase, whose mean is centered at $\varphi_k$, but whose values are drawn from an interval of the length of the difference of measured phases. With this Monte-Carlo-like approach, we efficiently calculated the nonclassicality quasiprobabilities.

\end{document}